# Plasma Scattering Measurement Using a Submillimeter Wave Gyrotron as a Radiation Source

I.Ogawa[1], T. Idehara[1], Y.Itakura[1], M.Myodo[1], T.Hori[2] and T.Hatae[3]

[1]*Research Center for Development of Far-Infrared Region, University of Fukui*
*3-9-1 Bunkyo, Fukui 910-8507, Japan*
[2]*Basic and Advanced Research Division, National Institute of Information and Communications Technology,*
*4-2-1, Nukui-Kita, Koganei, 184-8795, Japan*
3*Japan Atomic Energy Research Institute,*
*801-1 Mukoyama, Naka 311-0193, Japan*

**Abstract.** Plasma scattering measurement is effective technique to observe low frequency density fluctuations excited in plasma. The spatial and wave number resolutions and the S/N ratio of measurement depend on the wavelength range, the size and the intensity of a probe beam. A well-collimated, submillimeter wave beam is suitable for improving the spatial and wave number resolutions. Application of high frequency gyrotron is effective in improving the S/N ratio of the measurement because of its capacity to deliver high power. Unlike the molecular vapor lasers, the gyrotrons generate diverging beam of radiation with TEmn mode structure. It is therefore necessary to convert the output radiation into a Gaussian beam. A quasi-optical antenna is a suitable element for the conversion system under consideration since it is applicable to several $TE_{0n}$ and $TE_{1n}$ modes. In order to apply the gyrotron to plasma scattering measurement, we have stabilized the output ($P$=110W, $f$=354GHz) of gyrotron up to the level ($\Delta P/P$<1 %, $\Delta f$<10 kHz). The gyrotron output can be stabilized by decreasing the fluctuation of the cathode potential.

## I. INTRODUCTION

Density fluctuations in magnetically confined plasmas are important physical quantities to be measured for the basic study of plasma confinement, since the density fluctuations are theoretically expected to enhance the energy loss of the confined plasma across the magnetic field. The scattering method with electromagnetic wave makes it possible to observe frequency and wavenumber of the fluctuations directly with a spatial resolution.

Plasma scattering measurement is effective technique to observe low frequency density fluctuations excited in plasma. The performance of plasma scattering measurement (S/N ratio, the spatial and wave number resolutions) depends on intensity, wavelength range and quality of a probe beam. A submillimeter wavelength beam is most suitable for improving spatial and wave number resolutions.

Up to now, molecular vapor lasers and backward-wave oscillators have been used as the principal radiation sources in submillimeter wavelength range. However, their output powers are not so high. Application of high frequency gyrotron is effective in improving the S/N ratio of the measurement because of its capacity to deliver high powers [1, 2]. Unlike a molecular vapor laser, a gyrotron generates spreading radiation with TEmn mode structure. It is therefore necessary to convert the output radiation into a Gaussian beam ($TEM_{00}$ mode), which is suitable for an effective transmission and a well-collimated probe beam.

A quasi-optical antenna is a suitable element for the conversion system under consideration since it is applicable to several $TE_{0n}$ and $TE_{1n}$ modes. It should be noted, however, that the far-field of the linearly-polarized beam produced by the antenna consists of side lobes and a main beam, which is similar to a bi-Gaussian beam. A Gaussian beam can be obtained by converting the main beam. A con-focal mirror system with different focal lengths in different directions is used for the conversion.

The stabilization of gyrotron output power and frequency are required to improve the performance of the measurement. They can be stabilized by decreasing the fluctuation of the cathode potential. The fluctuation will be removed by using a smoothing circuits consisting of capacitor, coil and resistor.

A final goal is to produce an optimal probe beam (high quality, intense submillimeter wave beam) by using a new transmission line and to apply it to plasma scattering measurements with the aim of optimal performance of the measurement.

## II. Application of Submillimeter Wave Gyrotron (Gyrotron FU II) to Plasma Scattering Measurement

The gyrotron FU II is one of high frequency, medium power gyrotrons included in Gyrotron FU Series developed in Fukui University. In the gyrotron, as well as other gyrotrons included in the series, a narrow resonant cavity with high Q value is installed for achieving high separation between the cavity modes. Such a situation is important for high harmonic operation of high frequency gyrotrons. Because of this narrow cavity, our gyrotrons could be operated in many single modes at the fundamental and the second and third harmonics of electron cyclotron resonance. The Gyrotron FU II consists of an 8T super conducting magnet, water-cooled gun coils and sealed-off gyrotron tube. The electromagnetic wave generated in the cavity transmits in a circular wave guide and emitted from the vacuum window. We used TE161 mode operation at the second harmonic ($n=2$) resonance for plasma scattering measurement. The frequency is 354GHz (the corresponding wavelength is 0.85mm). For the application, we tried long pulse operation. Fig.1 shows the results. The pulse width is about 600ms, the output power is 110W and the electron beam energy and current are 28keV and 240mA, respectively. This pulse length and the output power are both enough for plasma scattering measurement.

In order to observe low frequency density fluctuations in the CHS (Compact Helical System) plasma, a plasma scattering measurement system with a gyrotron as a power source is installed (Fig.2). The gyrotron output is transmitted to a quasi-optical antenna through circular wave guides and two miter bends and then converted into a linearly-polarized beam. It is injected into the plasma as O-mode. The quasi-optical antenna plays two important roles. One is the mode conversion of the circular wave guide mode into a linearly-polarized mode and the other the focusing of the gyrotron output. The waves scattered by plasma density fluctuations are received by horn antennae installed in the plasma vessel and are converted into low frequency signals by a homodyne detection system. A compact Schottky barrier diode mounted in a corner cube is used as the mixer.

The frequency spectrum of the signal is obtained by FFT. The scattering measurements with scattering angles of $4.4°$ and $8.8°$ are carried out for ICRF ($f=26$MHz, $P=250$kW) heated plasmas. The scattering angle is determined by the configuration of the probe beam and the scattered wave (horn antennae). Target plasmas are produced by electron cyclotron resonance heating ($f=53.2$GHz, $P=200$kW, $t=13\sim43$ms) and heated further during the ICRF pulse ($t=40\sim90$ms). After the ICRF pulse, the energy stored in the plasma abruptly decreases and the plasma density becomes zero around $t=150$ms. Figure 3(a) shows time evolutions of scattered wave power for respective frequency intervals. This scattering angle of $8.8°$ corresponds to wave number of $11.4$cm$^{-1}$. The increase in scattered wave power is observed in the ICRF heating phase ($t=40\sim90$ms). However, no significant increase in scattered power is observed in high frequency range ($f>50$kHz). Reflectmetry at the frequency of 39GHz gives similar results (Fig.3(b)). This measurement system gives a S/N ratio high enough in the higher frequency range.

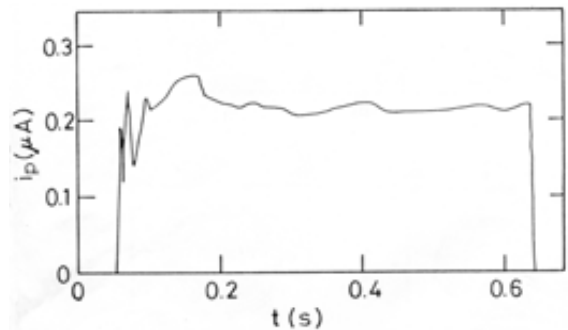

Fig.1 Time evolution of output power measured by a pyro-electric detector.

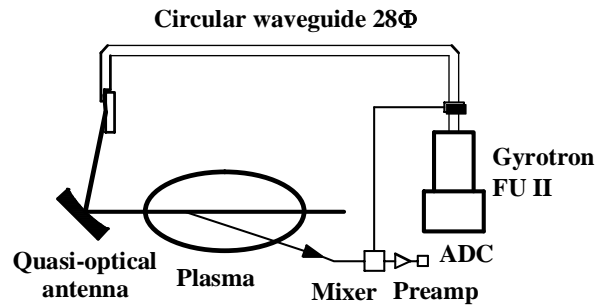

Fig.2 Block diagram of the submillimeter wave scattering measurement.

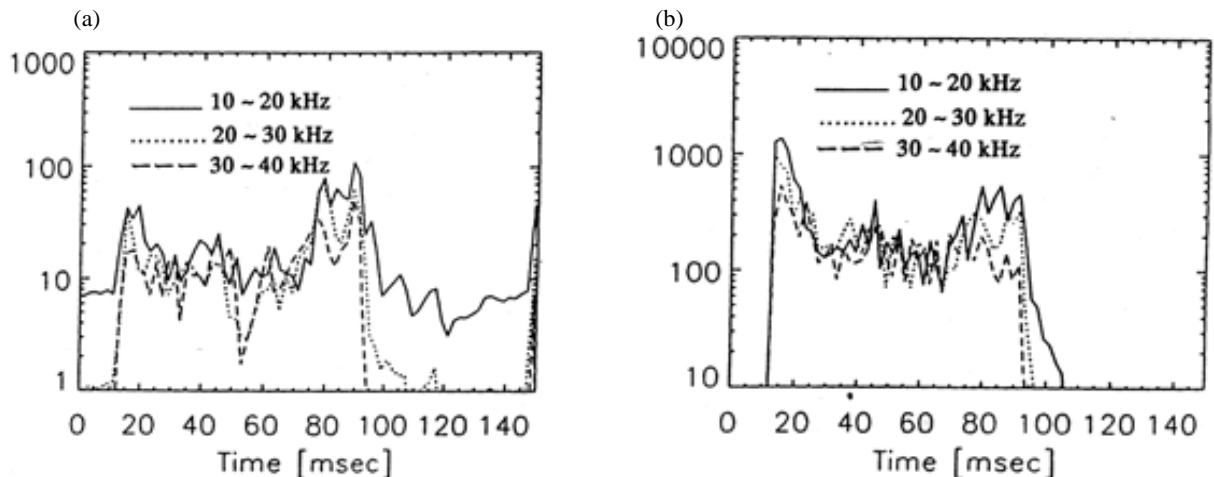

Fig.3 Time evolution of (a) scattered wave power and (b) signal of reflectmetery at the frequency of 39 GHz.

## III. Generation of Well-Collimated Beam

Unlike the molecular vapor lasers, the gyrotrons generate diverging beam of radiation with TEmn mode structure. It is therefore necessary to convert the output radiation into a Gaussian beam ($TEM_{00}$ mode), which is suitable for an effective transmission and can be used as a well-collimated probe beam. The system consists of two components, namely a quasi-optical antenna and an ellipsoidal mirror. The first element in the quasi-optical system is the quasi-optical antenna (Fig.4) which consists of stepped-cut launcher with radius 14mm and length of the step cut 220mm and a parabolic reflector whose focal length is 21.75mm. The antenna is followed by an ellipsoidal mirror (Fig.5). The antenna converts the gyrotron radiation of $TE_{0n}$ and $TE_{1n}$ operating modes into linearly-polarized beams and the ellipsoidal mirror produces the required far-field pattern. The measurements of the final beam structure have demonstrated that the system developed produces bi-Gaussian beams from the outputs of $TE_{0n}$ and $TE_{1n}$ operating modes.

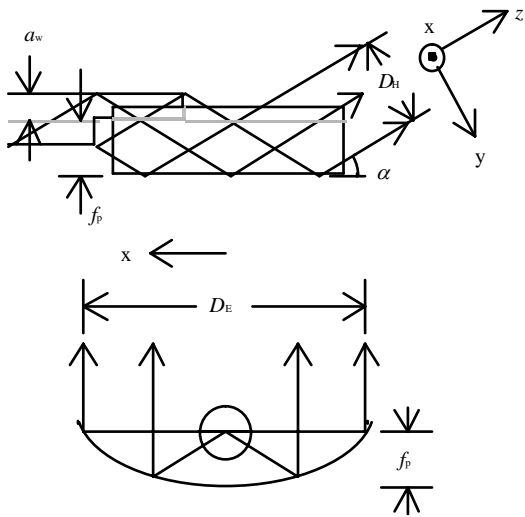
Fig.4 Quasi-optical antenna.

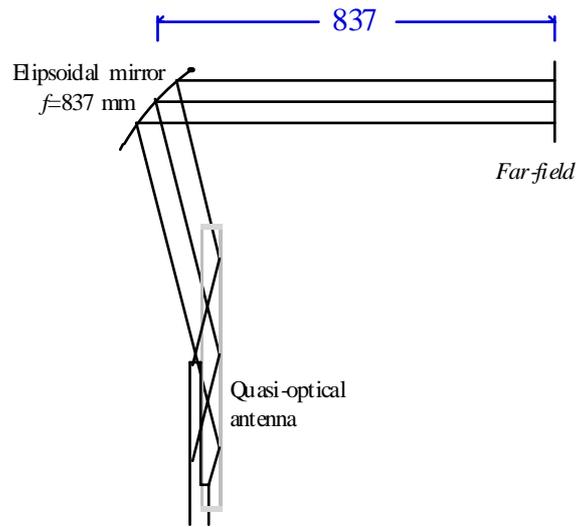
Fig.5 Quasi-optical system to convert gyrotron output into Gaussian beam.

## IV. Stabilization of Gyrotron Output and Frequency

We have achieved long pulse operation of the submillimeter wave gyrotron, Gyrotron FU II. The output power of long pulse operation was not so stable ($\Delta P/P \sim 10\%$) due to the fluctuation of the cathode potential ($\Delta V_k \sim 40V$). The evolution of operation parameters with time is shown in Fig.6. The fluctuation of the output power correlates with that of the cathode potential $\Delta V_k$. In order to suppress the fluctuation level of the cathode potential, high voltage power supplies are equipped with smoothing circuits consisting of a resistor, an induction coil and a capacitor (Fig.7). The fluctuation level was decreased ($\Delta V_k \sim \sim 1V$) by introducing the smoothing circuit. Accordingly, the fluctuations of the output power were decreased from 10% to 1% (Fig.8) and those of frequency was decreased from 100kHz to 10kHz (Fig.9).

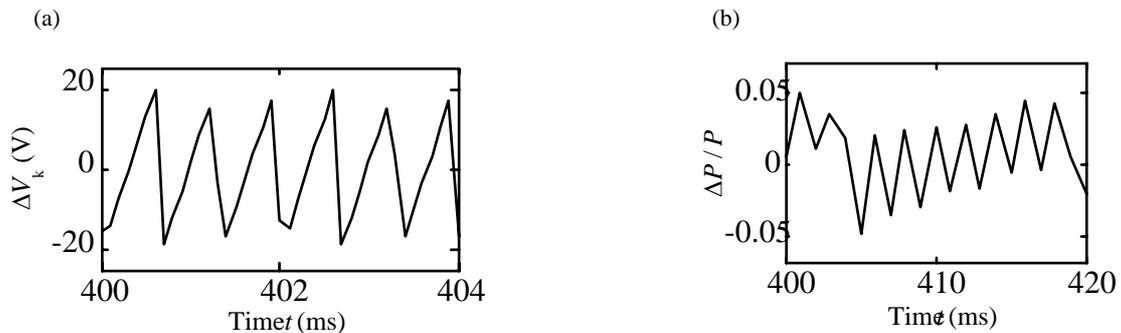
Fig.6 Time evolutions of cathode voltage fluctuation (a) without a smoothing crcuit and (b) that with a smoothing circuit.

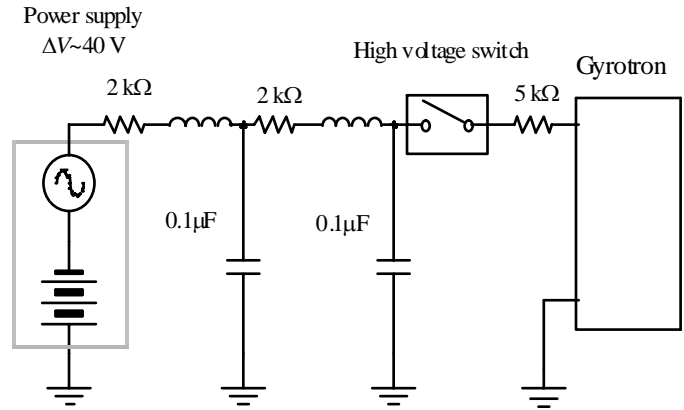

Fig.7 Smoothing circuit to decrease fluctuation of cthode voltage.

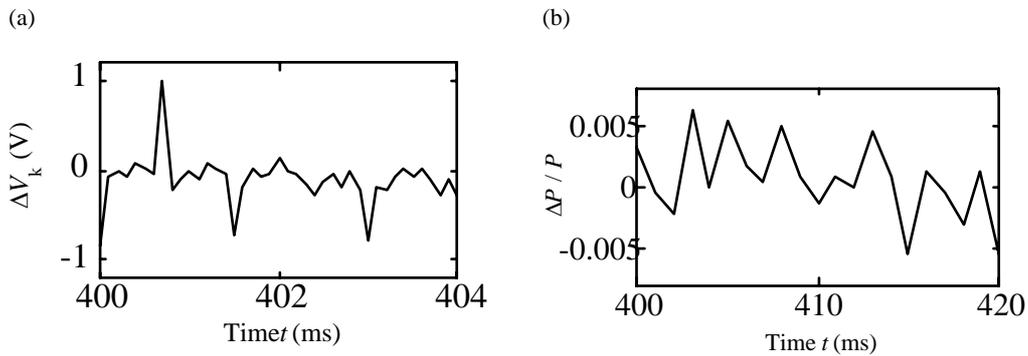

Fig.8 Time evolutions of output power fluctuation (a) without a smoothing crcuit and (b) that with a smoothing circuit.

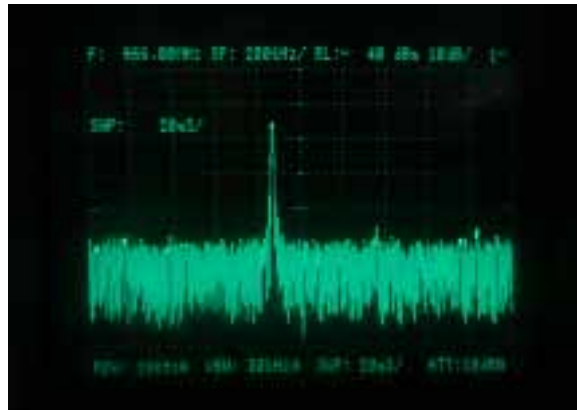

Fig.9 Frequency spectrum of intermediate frequency signal obtained by heterodyne detection system.

## V. Conclusions

The GYROTRON FU II delivers long pulses of suitably high power in submillimeter wavelength range ($P$~110W, $f$=354GHz). Output power radiation has been applied to scattering measurements in the CHS plasma. The output of the Gyrotron FU II obtained by long pulse operation is stabilized up to 1% and 10kHz by decreasing the fluctuation of the cathode voltage ($\Delta V_k$~1V), via introducing a smoothing circuit consisting of a resistor, an induction coil and a capacitor.